\documentclass[%
 reprint,
 amsmath,amssymb,
 aps,
prb,
superscriptaddress
]{revtex4-2}

\usepackage[dvipsnames]{xcolor}
\usepackage{braket}
\usepackage{graphicx}
\usepackage{dcolumn}
\usepackage{bm}
\usepackage{float}
\usepackage{physics}
\usepackage{verbatim}
\usepackage{hyperref}
\hypersetup{
    colorlinks,%
    citecolor=blue,%
    linkcolor=blue,%
    urlcolor=blue
}

\usepackage{soul}

\begin{document}


\title{Near zero-energy Caroli-de Gennes-Matricon vortex states in the presence of impurities}

\author{Bruna S. de Mendon\c{c}a}
\email{bsmend@usp.br}
\affiliation{
Instituto de F\'{\i}sica, Universidade de S\~ao Paulo, Rua do Mat\~ao 1371, S\~ao Paulo, S\~ao Paulo 05508-090, Brazil
}

\author{Antonio L. R. Manesco}
\affiliation{
Kavli Institute of Nanoscience, Delft University of Technology, Delft 2600 GA, The Netherlands}%

\author{Nancy Sandler}
\affiliation{
Department of Physics and Astronomy, Ohio University, Athens, Ohio 45701, USA}%

\author{Luis G. G. V. Dias da Silva}
\affiliation{
Instituto de F\'{\i}sica, Universidade de S\~ao Paulo, Rua do Mat\~ao 1371, S\~ao Paulo, S\~ao Paulo 05508-090, Brazil
}

\date{\today}

\begin{abstract}
 Caroli-de Gennes-Matricon (CdGM) states are localized states with a discrete energy spectrum  bound to the core of vortices in superconductors.
In topological superconductors, CdGM states are predicted to coexist with zero-energy, chargeless states widely known as Majorana zero modes (MZMs).
Due to their energy difference, current experiments rely on scanning tunneling spectroscopy methods to distinguish between them.
This work shows that electrostatic inhomogeneities can push trivial CdGM states arbitrarily close to zero energy in non-topological systems where no MZM is present.
Furthermore, the BCS charge of CdGM states  is suppressed under the same mechanism.
Through exploration of the impurity parameter space, we establish that these two phenomena generally happen in consonance.
Our results show that energy and charge shifts in CdGM may be enough to imitate the spectroscopic signatures of MZMs even in cases where the estimated CdGM level spacing (in the absence of impurities) is much larger than the typical experimental level broadening.
\end{abstract}

\maketitle
\section{ Introduction } \label{sec:Intro}

Andreev bound states are a class of low-energy quasiparticle excitations that appear in metallic regions confined by a superconducting gap.
A particular type of Andreev bound state excitations exists on zero-dimensional defects in topological superconductors.
These zero-energy excitations are known as Majorana zero modes (MZMs)~\cite{beenakker2019search,sarma2015majorana,fu2008superconducting,sau2010generic,kong2021emergent}.
Since these quasiparticles are topologically-protected non-Abelian anyons~\cite{KITAEV20032,AguadoReview,alicea2012new}, they can be used to build fault-tolerant qubits~\cite{RevModPhys.80.1083}.
The interest in building robust quantum computers led to intense efforts to search for and identify MZMs.

\begin{figure}[!ht]
    \centering
    \includegraphics[width=\columnwidth]{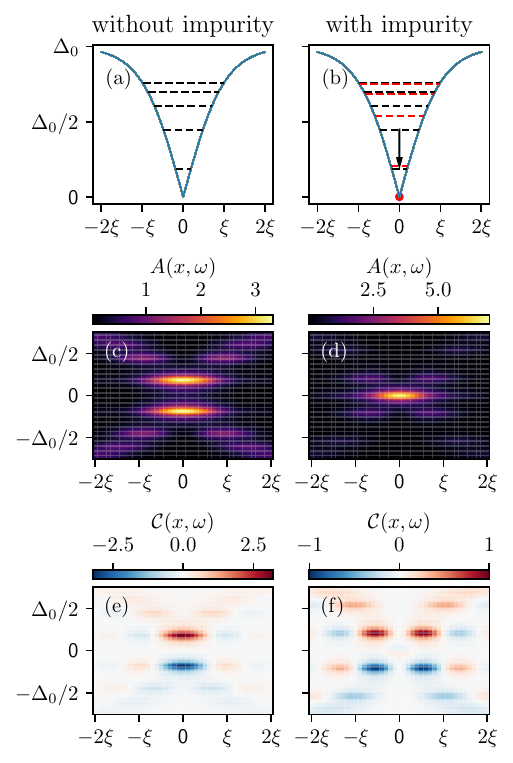}
    \caption{
    Energy states inside the vortex for (a) a clean $s$-wave superconductor and (b) an $s$-wave system with screened charge impurity.
    The blue line shows the position-dependent superconducting order parameter, and the dashed lines schematically show the energy of in-gap CdGM states.
    In panel (b), the black dashed lines indicate the CdGM spectrum without a charged impurity, whereas the red  dashed lines indicate the spectrum with the impurity.
    The black arrow highlights the energy shift caused by the impurity potential.
    Panels (c) and (d) show the density of states, whereas (e) and (f) show the BCS spectral charge for the $s$-wave system without and with impurity respectively.
    }
    \label{fig:fig1}
\end{figure}

The first generation of experiments to identify Majorana modes relied on measuring zero-bias peaks with tunneling probes.
These peaks indicate the existence of zero-energy excitations.
However, to interpret these zero-bias peaks as Majorana zero modes, one must rule out the existence of other (trivial) excitations.
Due to this ambiguity, together with theoretical progress showing that zero-energy trivial states are possible, a new generation of experiments combining local and nonlocal probes was recently developed in nanodevice platforms~\cite{pan2020physical, rosdahl2018andreev, zhang2019next, menard2020conductance, danon2020nonlocal,Cao:Phys.Rev.B:075416:2022,Kouwenhoven_arxiv221106709,pikulin2021protocol,aghaee2022inas}.

In bulk topological superconductor candidates, an alternative procedure to detect Majorana zero modes is by piercing the superconductor with a magnetic field.
With an applied field, superconducting vortices are formed.
At sufficiently low electron density, these vortices should host Majorana zero modes isolated from other in-gap states.
Thus, one can again probe the MZMs by scanning tunneling spectroscopy.
Again, one of the fundamental challenges in verifying the presence of MZMs in vortices of topological superconductor candidates is establishing a clear distinction from other trivial states.

In superconducting vortices, non-topological quasiparticle excitations are known as Caroli-de Gennes-Matricon (CdGM) states~\cite{prada2020andreev}.
They appear in low-density superconducting materials, where vortices act as ``quantum wells for quasiparticles''~\cite{abrikosov2004Nobel}.
In trivial superconductors, vortices contain bound states with a low-energy spectrum given by \cite{caroli1964bound} (see Fig.~\ref{fig:fig1}~(a, b)):
\begin{equation}
    \label{eq:cdgm_energy}
    E_m = \frac{m \Delta_0}{k_F\xi}~,
\end{equation}
where $m = (n +1/2)$, $n$ is an integer number, $k_F$ is the Fermi momentum, $\xi$ the bulk coherence length, and $\Delta_0$ is the bulk superconducting pairing potential.
The predicted spectrum of CdGM states in trivial superconductors lacks a zero-energy level.
These features establish the difference between CdGM and MZMs spectra and underlie the interpretation of experimental observations of zero-bias peaks inside vortices~\cite{xu2015experimental,kong2021majorana,hou2021zero, kong2021emergent, PhysRevX.9.011033, machida2019zero, 2021ApPRv...8c1417K, wang2021spinpolarized, wang2018evidence,Chen:NatureCommunications:970:2018}.
From Eq.~\eqref{eq:cdgm_energy}, it is clear that an energy resolution better than $\Delta_0 /k_F\xi$ is required to detect isolated excited levels.
Since this resolution is accessible in state-of-the-art experiments, tunneling spectroscopy measurements of vortices on topological superconductors are expected to distinguish Majorana from trivial states~\cite{sbierski2021identifying, chiu2020scalable, PhysRevB.104.L020505}.

Recent experiments on candidate topological superconductor materials revealed that many zero-bias peaks, associated with the presence of MZM modes, were not present in all vortices \cite{Chen:NatureCommunications:970:2018,machida2019zero}.
Furthermore, some of the detected peaks often appeared to be stabilized by nearby magnetic~\cite{wang2021spinpolarized} or scalar~\cite{kong2021majorana} impurities.
Although it is well known that zero-bias peaks can emerge in superconductors without topological properties when magnetic impurities are present, it is still an open question whether scalar impurities may produce similar effects~\cite{PhysRevLett.126.257002}.

Motivated by this scenario, we study the effect of scalar impurities in the in-gap spectrum of a two-dimensional trivial $s$-wave superconductor.
We revisit this system because: (i) $s$-wave superconductor is a simple example of a trivial superconductor;
(ii) the Hamiltonian does not contain additional terms that could lead to corrections from Eq.~\eqref{eq:cdgm_energy}.
As a result,  the spectra and corresponding charge distributions can be traced back unequivocally to the presence of the impurity potential.

Moreover, influenced by the new generation of experiments in nanowires, 
non-local measurements with scanning tunneling microscope measurements were suggested as a way to extract the Bardeen-Cooper-Schrieffer (BCS) charge of in-gap vortex excitations \cite{sbierski2021identifying}.
Because trivial states have non-zero BCS charge whereas MZMs are strictly chargeless~\cite{AguadoReview}, information on the BCS charge in principle helps to distinguish CdGM states and MZMs.

Our main results are illustrated in Fig.~\ref{fig:fig1}, which compares the density of states and BSC charge for a trivial $s$-wave superconductor with and without an impurity potential.
The calculations reveal that trivial CdGM states mimic MZM's signatures in two ways.
First, electrostatic inhomogeneities shift the lowest CdGM state energy arbitrarily close to zero, as schematically shown in Figs.~\ref{fig:fig1}~(a, b).
Second, as the energy separation of these states become smaller than the experimental resolution -- mainly limited by thermal broadening --, the spectral BCS charge is also suppressed, as shown in Figs.~\ref{fig:fig1}~
(e, f).
As a consequence, electrostatic inhomogeneities can make CdGM indistinguishable from MZMs when probed by STM experiments, given the current experimental resolutions.

\section{Model}
\label{sec:model}

\begin{figure}[!t]
    \centering
    \includegraphics[width=1\columnwidth]{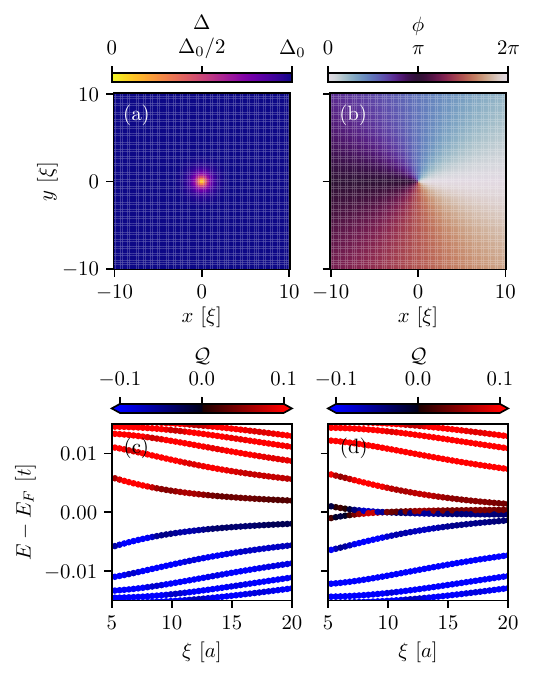}
    \caption{
    Magnitude (a) and phase (b) of the superconducting order parameter (as defined in Eq. \eqref{eq:Deltaswave}).
    The BCS charge-resolved spectrum of in-gap vortex states is shown for $s$-wave superconductor (c) without and (d) with an screened charge impurity with
    $\delta \mu = 0.06t$ and $\eta=5a$.}
    \label{fig:fig2}
\end{figure}

We consider a trivial superconductor modeled as a two-dimensional electron gas with $s$-wave superconducting pairing.
The corresponding tight-binding model on a square lattice (with lattice constant $a$) has a normal state Hamiltonian
\begin{equation}
    H_0 = (4t -\mu) \sum_{i} \sum_{\sigma}c_{i \sigma}^{\dagger} c_{i \sigma} -t \sum_{\langle i, j\rangle}\sum_{\sigma}c_{i \sigma}^{\dagger} c_{j \sigma}~,
\end{equation}
where $\mu$ is the onsite energy, $t$ is the hopping constant, $c_{i \sigma} ^{\dagger} (c_{i \sigma})$ creates (destroys) an electron of spin $\sigma$ 
at site $i$, and $\langle i, j\rangle$ denotes a sum performed over nearest-neighbors.
The superconducting term in the Hamiltonian is
\begin{align}
    H_{s\text{-wave}} &= \sum_{i} \Delta_i^s c_{i \uparrow}^{\dagger} c_{i \downarrow}^{\dagger} + h.c.,  \nonumber \\ 
    \Delta_i^s &= \Delta_0 e^{i \phi_i} \tanh(\frac{r_i}{\xi})~, \label{eq:Deltaswave}
\end{align}
where $\Delta_0$ is the amplitude of the bulk order parameter, $r_i$ is the distance from $\mathbf{r}=0$ (the vortex center) to the atomic position $i$, $\xi$ is the vortex radius, and $\phi_i=\arg(\mathbf{r}_i)$ is the order parameter phase, as shown in Figs.~\ref{fig:fig2}~(a) and (b).
Note that we treat $\xi$ and $\Delta_0$ as independent parameters, since we do not solve the Ginzburg-Landau equations.
To match the typical ratios for $\mu/ \Delta_0$ reported in experiments we set $\mu=0.05t$ and $\Delta_0=0.02t$.
We also set $\xi=10a$, unless stated otherwise, to ensure that the vortex size is negligible compared to the system size ($200a \times 200a$).
We treat spin as a trivial degeneracy and perform all tight-binding calculations with Kwant~\cite{Groth_2014}.

Figure~\ref{fig:fig2}~(c) show the spectrum and the respective integrated BCS charge expectation values, $\mathcal{Q} \equiv \sum_i \mathcal{Q}(\mathbf{r}_i)$ where $\mathcal{Q}(\mathbf{r}_i) := \langle \Psi | c_i^{\dagger}c_i - c_i c_i^{\dagger} | \Psi \rangle$ is the local charge for a given eigenstate, of in-gap vortex states for this system as a function of the vortex radius $\xi$.
Finite energy levels approach zero as $\xi$ increases, as expected for quantum-confined levels.

\section{Effects of an impurity potential }
\label{sec:impurity}

An inspection of Eq.~\eqref{eq:cdgm_energy} suggests that local changes in the electrostatic potential can shift the energy of CdGM states arbitrarily close to zero energy.
In this situation, the lowest-energy trivial states may have energy smaller than the experimental resolution, resulting in a near-zero-bias conductance peak, similar to the one produced by MZMs.
To demonstrate this phenomenon, we consider the effects of an isolated screened charge impurity close to the vortex core.
This choice is inspired by recent experiments in iron-based superconductors~\cite{Chen:NatureCommunications:970:2018,kong2021majorana,machida2019zero} showing that scalar impurities favor the presence of zero-bias peaks inside vortices.

We model the presence of a screened charged impurity on the $s$-wave Hamiltonian by incorporating an onsite modulation term as follows \cite{rycerz2007anomalously, wurm2012symmetries}:
\begin{equation}\label{eq:impurity}
    H_{\rm imp} = \delta \mu \sum_{i, \sigma} \ e^{-r_i^2 /2\eta^2}c_{i \sigma}^{\dagger}c_{i \sigma}~,
\end{equation}
where $\delta \mu$ is the potential strength of the impurity, and $\eta$ is the screening length.
Next, we calculate the energy spectra and the integrated charge $\mathcal{Q} \equiv \sum_i \mathcal{Q}(\mathbf{r}_i)$ of the Andreev quasiparticles.

\begin{figure}[t]
    \centering
    \includegraphics[width=\columnwidth]{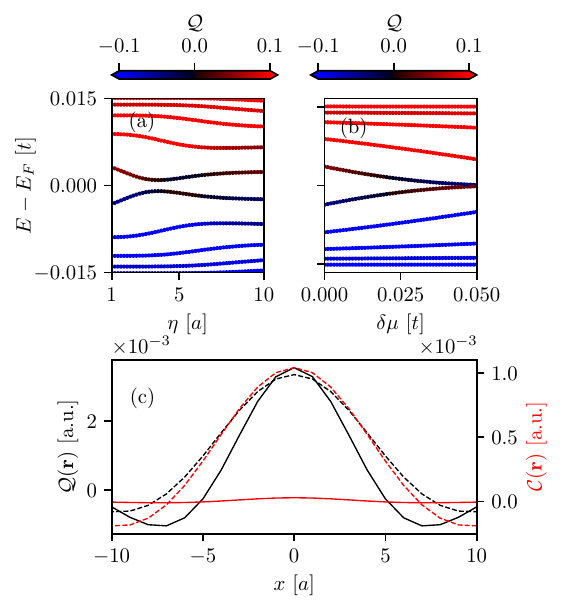}
    \caption{
    BCS charge for an s-wave system with impurity.
    In panels (a,b), the spectrum evolves as a function of (a) impurity size $\eta$ ($\delta\mu = 0.025 t$),
    (b) impurity strength $\delta \mu$ ($\eta=5a$).
    (c) BCS charge (black) and BCS charge spectral density (red) for the lowest energy state for a system without 
    (dashed line) and with (solid line) an impurity with $\eta=5a$  
    and $\delta\mu = 0.05t$.
    }
    \label{fig:fig3}
\end{figure}

Figure \ref{fig:fig3} shows the resulting low-energy spectra as a function of $\eta$ and $\delta \mu$, revealing a resemblance to the spectra of a topological superconductor with a vortex.
Furthermore, both charge and energy of the lowest-energy state approach zero, while the shifts of energy and charges of higher energy states are exponentially suppressed 
(see Appendix \ref{app:perturbation} 
for a quantitative discussion).

The local BCS charges $\mathcal{Q}(\mathbf{r})$ for the lowest-energy in-vortex state with and without the impurity are shown in Fig.~\ref{fig:fig3}~(c).
We verified that, although the total charge is suppressed [Fig.~\ref{fig:fig3}(a, b)], the local charge is weakly affected by the impurity.
Thus, one should be able to distinguish CdGM and Majorana states with an arbitrarily small resolution, since MZMs have zero BCS charge everywhere.
However, as we show in the next section, the energy shift together with level broadening of the lowest-energy CdGM states results in a vanishing spectral BCS charge.

\begin{figure}[!t]
    \centering
    \includegraphics[width=1\columnwidth]{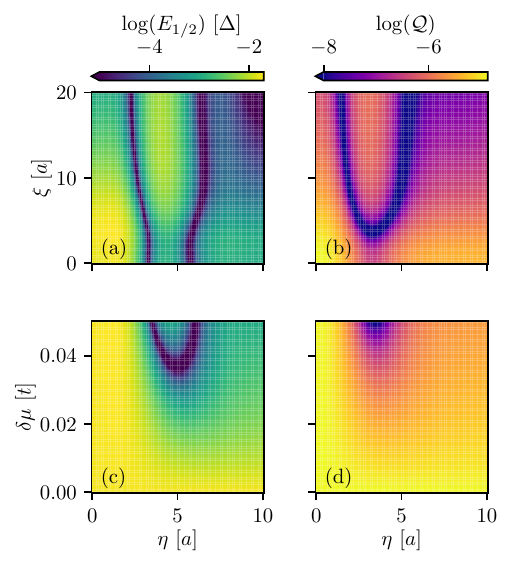}
    \caption{Dependence of lowest energy state $E_{1/2}$ (a, c) and BCS charge (b, d) on (a, b) $\eta$ and $\xi$, (c, d) $\eta$ and $\delta \mu$.
    In (a, b) we choose $\delta \mu = 0.06 t$, and in (c, d) we choose $\xi = 5 a$.
    }
    \label{fig:fig4}
\end{figure}

The effects of the vortex size and impurity screening length on the energy and charge of the lowest-energy state ($E_{1/2}$, as defined in Eq.~\eqref{eq:cdgm_energy})  are shown in Fig.~\ref{fig:fig4}.
The data clearly shows a suppression of $E_{1/2}$ and $\mathcal{Q}$ on large regions of the $\{ \eta$, $\delta \mu \}$ parameter space, indicating that the above results hold beyond the highly-localized impurity regime, $\eta \ll \xi$. This also suggests that smooth fluctuations in the underlying electrostatic potential can mimic Majorana signatures.
Finally, let us note that, due to particle-hole symmetry, $E_{1/2}$ and $\mathcal{Q}$ tend to be correlated.

\section{Experimental relevance} 
\label{sec:experiments}

In this Section, we aim to establish the relevance of our results for the interpretation of current experimental data from  Scanning Tunneling Spectroscopy (STS) and Scanning Tunneling Microscopy (STM) experiments \cite{wang2018evidence,Chen:NatureCommunications:970:2018,machida2019zero}. 

Thus far, we have discussed the changes in the global properties of CdGM state due to the the presence of scalar impurities at the vortex site. Since vortices tend to be pinned by defects and impurities \cite{kreisel2020remarkable}, this picture can be favored in samples in which some degree of surface disorder is present, as it seems to be the case in Fe(Se,Te) surfaces \cite{machida2019zero}. 

Moreover, experimental estimates of $E_{1/2}$ (as defined in Eq.~\eqref{eq:cdgm_energy}) are commonly used as a proxy for the position of the first CdGM state. As such, STS conductance peaks at energies below this estimate are usually ``ruled out'' as CdGM states \cite{wang2018evidence,machida2019zero}.
As we have shown, such a heuristic picture is not accurate in the presence of impurities: the results shown in Figs. \ref{fig:fig2} and \ref{fig:fig3} show that the energy of the first CdGM state can be significantly lower than the value of $E_{1/2}$ estimated from bulk parameters and serve as a cautionary tale against ruling out these near-zero states as CdGM states.

Another important point of attention when comparing the raw CdGM spectra with the peaks appearing in STS measurements is the role of the thermal effects of the STS peaks, which effectively sets an energy resolution. To illustrate the limitations introduced by such energy resolution, we defined the spatially-resolved  spectral  density $A(\omega, \mathbf{r})$,  and the spectral BCS charge $\mathcal{C}(\omega, \mathbf{r})$ at energy $\omega$ and position $\mathbf{r}$: 
\begin{align}
    G(\omega) &= [\omega - H + i \gamma]^{-1}~, \\[0.1cm]
    G(\omega, \mathbf{r}) &\equiv \langle \mathbf{r} | G(\omega) |  \mathbf{r} \rangle~,\\  
    A(\omega, \mathbf{r}) &\equiv -\dfrac{1}{\pi} \Im \Tr \left[G(\omega, \mathbf{r})\right]~, \\
    \mathcal{C}(\omega, \mathbf{r}) &= -\dfrac{1}{\pi} \Im \Tr \left[G(\omega, \mathbf{r}) \mathcal{Q}(\mathbf{r})\right]~,   
\end{align}
where $\gamma$ is a positive small parameter that sets the level broadening.

\begin{figure}[!t]
    \centering
    \includegraphics[width=1\columnwidth]{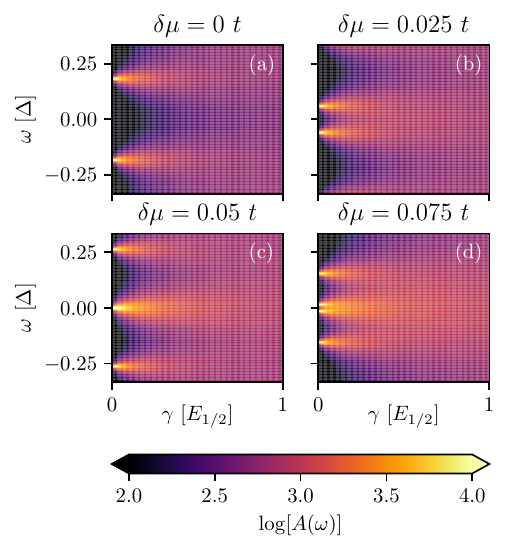}
    \caption{
        Density of states $A(\omega) \equiv (-1/\pi) \mbox{Im Tr  }G(\omega)$ of an $s$-wave superconductor as a function of level broadening $\gamma$ ($\eta = 5a$).
        In panel (a), one can observe that the two trivial states are distinguishable in the absence of an impurity ($\delta \mu =0$) up to $\gamma \sim E_{1/2}$.
        As the impurity strength increases, as shown in panels (b-d), the trivial states are shifted to smaller energies $\omega$ and therefore one cannot distinguish them even with a much smaller level broadening $\gamma$. 
    }
    \label{fig:fig5}
\end{figure}

Typical experimental broadenings are of order $\gamma \sim 0.1 \; \Delta_0$ (mostly arising from thermal effects) whereas the energy spacing estimated without impurities is $E_{1/2} \sim 0.5 \Delta_0$. \cite{wang2018evidence}
Thus, for clean systems, scanning tunneling transport experiments should indeed be able to resolve the energies of MZMs and CdGM states.
However, electrostatic inhomogeneities that make $E_{1/2} \sim \gamma$ impede a clear distinction between trivial and non-trivial states due to the constraint imposed by particle-hole symmetry that implies the shifting of positive and negative energy states towards zero energy.

We illustrate the effects of level broadening in Figs.~\ref{fig:fig1} and~\ref{fig:fig5}.
Figure \ref{fig:fig1} shows contour plots of $A(\omega, \mathbf{r})$ [Fig.~\ref{fig:fig1}(c,d)] and $\mathcal{C}(\omega, \mathbf{r})$ [Fig.~\ref{fig:fig1}(e,f)] for energies within the gap and positions inside the vortex.
In the absence of impurities [Fig.~\ref{fig:fig1}(c) and (e)], one obtains peaks in $A(\omega  \approx \pm E_{1/2}, \mathbf{r} \approx 0)$ originating from the first CdGM state, as expected.
Some smaller peaks for excited CdGM states (with nodes at $\mathbf{r} \approx 0$) are also seen.
The BCS charge for these states oscillates in position along the vortex, changing sign.

A different picture is observed in the presence of impurities and a modest thermal broadening ($\gamma/E_{1/2} \approx 0.27$).
There is a single peak in the spectra charge density near zero energy $A(\omega \approx 0, \mathbf{r} \approx 0)$, and the BCS charge at these energies is very close to zero.
In Fig.~\ref{fig:fig5} we illustrate how much the level broadening affects our results.
In the absence of impurities [Fig.~\ref{fig:fig5}(a)], the energies of CdGM states are determined by the bulk parameters.
Therefore, the two lowest-energy peaks are distinguishable up to $\gamma/E_{1/2} \sim 1$.
As the energy levels shift towards zero, 
the two lowest-energy peaks cannot be resolved for $\gamma/E_{1/2} \gtrsim 0.5$ (Fig.~\ref{fig:fig5}~(b)). In some cases, these peaks cannot be resolved even for relatively ``small" broadening values $\gamma/E_{1/2} \sim 0.1$ or lower (Fig.~\ref{fig:fig5}~(c-d)).
In a similar manner, since particle-hole symmetry ensures that $\mathcal{Q}_{m=1}(\mathbf{r}) = - \mathcal{Q}_{m=-1}(\mathbf{r})$, the BCS charge also becomes negligible, as shown in Fig.~\ref{fig:fig1}~(f).

Notice that the combined effect of the shifting of the energy levels toward zero and a moderate broadening can effectively ``disguise'' the lowest CdGM state as a ``zero-bias peak'' (see. e.g., Figs. \ref{fig:fig5} (c,d)) in experiments.
Interestingly, some STS experimental data show vortex bound states as peaks \emph{near but not at} zero bias in the tunneling spectra \cite{Chen:NatureCommunications:970:2018}, a result which is consistent with this picture.

These observations suggest that recent experiments claiming impurity-assisted formation of MZMs \cite{kong2021majorana} should be interpreted carefully.
At the same time that local changes in the chemical potential could lead to a topological phase transition, trivial CdGM states can also be shifted arbitrarily close to zero energy.
Thus, zero-bias peaks in the vicinity of scalar impurities are ambiguous signatures and cannot fully distinguish MZM and CdGM states.

\section{Concluding Remarks}
\label{sec:Conclusions}

Our main results can be summarized as follows: the presence of scalar impurities in vortices can modify the CdGM spectrum, leading to similar spectral properties to MZMs.
We illustrate this fact in the extreme limiting case of a non-topological (s-wave) superconductor showing zero energy states and a vanishing global BCS charge.

This is, in fact, a very general result: local scalar perturbations in the vicinity of the vortex center lower the energy of CdGM states and suppress their total BCS charge. Particularly, the energy and charge of CdGM modes can be shifted arbitrarily close to zero for a broad range of values in the parameter space of the model proposed, as shown in Fig.~\ref{fig:fig4}.
As a consequence, such impurity-driven zero energy states cannot be set apart from topological MZMs using only local spectroscopic techniques.
Furthermore, due to extrinsic broadening of the particle-hole symmetric levels, near zero-energy states will render vanishing non-local transport signals, therefore resulting in the apparent vanishing of the BCS charge.
These considerations are particularly relevant to experimental works that often rely on the fact that CdGM states have finite energy.

One possible route to complement the current local STS set-ups could be provided by \emph{non-local} probes.
As argued in Ref. \cite{sbierski2021identifying}, non-local transport measurements could, in principle, map the changes in the BCS charge at a local level provided that the experimental resolution allows for a distinction between  MZM and CdGM states.
Thus, it is important to provide a description of the spectroscopic properties as they will be affected by the presence of electrostatic inhomogeneities.

\paragraph*{Acknowledgements.}
This study was financed in part by the Coordenação de Aperfeiçoamento de Pessoal de Nível Superior - Brasil (CAPES) - Finance Code 001.
The work of ALRM was sponsored by a NWO VIDI Grant (016.Vidi.189.180).
LGDS acknowledges financial support from Brazilian agencies FAPESP (Grant 2016/18495-4) and CNPq (Grants 423137/2018-2, and 309789/2020-6). Portions of this work were completed at NBI (KU) and the Physics Department at DTU (Denmark) under support from Otto M{\o}nsteds and NORDEA foundations (NS). 

\paragraph*{Data availability.}
All the scripts and resulting data used to prepare this manuscript are freely available on Zenodo \cite{bruna_s_de_mendonca_2022_6444338}.

\paragraph*{Authors' contributions.}
LGDS and NS formulated the initial project that was later refined with contributions from all authors.
BSM and ALRM carried out the numerical simulations and analyzed the data with input from the other authors. ALRM identified the role of scalar impurities on the in-gap spectrum. NS and LGDS supervised the project. BSM wrote the initial draft of the manuscript.
All authors contributed to writing the manuscript.

\onecolumngrid

\appendix

\section{Perturbative corrections due to the impurity potential}
\label{app:perturbation}

From Fig.~\ref{fig:fig3}~(b) it is evident that lower energy states have larger shifts than higher energy ones.
This quantitative difference can be traced back to the nature of the confined wavefunctions and is obtained using perturbation theory arguments.

For an $s$-wave superconductor, the eigenstates are written as \cite{khodaeva2022vortex}:
\begin{align}
    \Psi_m(r)
    = \left( \begin{array}{c}
    u_m(r) \\
    v_m(r)
    \end{array} \right)
    \propto e^{-K(r)}
    \left( \begin{array}{c}
    J_m(k_Fr) \\
    J_{m+1}(k_F r)
    \end{array} \right),
    \label{eq:spinor}
\end{align}
with $J_m(k_Fr)$ being the order-$n$ Bessel function of the first kind and
\begin{equation}
    K(r) = \frac{1}{\hbar v_F} \int_0^r dr'~ \Delta(r')~,
\end{equation}
where $v_F$ is the Fermi velocity in the normal state.
For low-energy in-gap states $E_n \ll \Delta_0$, we expand the superconducting pairing as $\Delta(r) \approx \frac{\Delta_0}{\xi}r$. Therefore:
\begin{equation}
    K(r) \approx \frac{\Delta_0}{2 \hbar v_F \xi}r^2=:\frac{r^2}{2\tilde{\xi}}~.
\end{equation}

The first-order corrections to the energy levels are given by
\begin{align}
    \delta E_m = \langle \Psi_m | H_{imp} | \Psi_m\rangle.
\end{align}
Assuming that there are not many states inside the vortex and particularly considering the low-energy ones, the Bessel functions can be approximated by their asymptotic form with $k_Fr \ll 1$.

In this regime, the energy correction follows a power law in $k_F^2/\alpha$, given by:
\begin{equation}
\delta E_m = \frac{\pi\delta \mu}{m!\alpha} \left(\frac{k_F^2}{4\alpha}\right)^{m}~,\quad \alpha:= \frac{1}{2\eta^2} + \frac{1}{2\tilde{\xi}^2}~.
\end{equation}

By carrying out first-order corrections on the eigenstates, we can understand the perturbative effects of the impurity on the charge. We find:
\begin{equation}
\label{eq:charge_correction}
\delta Q_m = 2\sum_{n\neq m} \frac{\beta_{mn}(\delta \mu, \eta)\beta_{mn}(1, \infty)}{E_m - E_n}.
\end{equation}
with
\begin{align}
\nonumber &\beta_{mn}(\delta \mu, \eta):=\langle \psi_m | \delta H | \psi_n \rangle \approx \frac{\pi \delta \mu}{m-n} \left[\frac{1}{m!n!} \left(\frac{k_F^2}{4\alpha}\right)^{\frac{m+n}{2}}\right] 
\frac{\Gamma((m+n+2)/2)}{\alpha}~.
\end{align}
Taking the lowest order term in Eq. \ref{eq:charge_correction}, we find:
\begin{align}
    \delta Q_m = 
    \begin{cases}
        \frac{\pi^2 \mu \delta \mu}{2\alpha^2\Delta_0^2}  \left(\frac{k_F^4\tilde{\xi}^2}{2\alpha}\right)^{\frac{3}{2}}\Gamma(5/2), & m=1\\
        \frac{\pi^2 \mu \delta \mu}{2\alpha^2 \Delta_0^2(m-1)^3} \left[\frac{1}{m!^2} \left(\frac{k_F^4\tilde{\xi}^2}{2\alpha}\right)^{\frac{m+1}{2}}\right] 
    \Gamma\left(\frac{m+3}{2}\right), & m\neq1
    \end{cases}~.
\end{align}
Therefore we obtain a polynomial dependence on $k_F^4\tilde{\xi}^2/2\alpha$. The results lead to the conclusion that the higher-energy states have smaller corrections in both energy and charge.

\twocolumngrid


%

\end{document}